\documentclass[12pt,preprint]{aastex}
\usepackage{graphicx}
\usepackage{txfonts}
\usepackage{natbib}
\bibpunct{(}{)}{;}{a}{}{,} 

\received{\ldots}
\accepted{\ldots}
\shortauthors{M. Adam\'ow et al.}
\begin{document}

  \title{BD+48 740 -- Li overabundant giant star with a planet. A case of recent engulfment?}

  \author{M. Adam\'ow$^1$, A. Niedzielski$^1$, E. Villaver$^2$, G. Nowak$^1$, A. Wolszczan$^{3,4}$}

  \altaffiltext{1}{Toru\'n Centre for Astronomy, Nicolaus Copernicus University, Gagarina 11, 87-100 Toru\'n, Poland}
  \altaffiltext{2}{Departamento de F\'{\i}sica Te\'orica, Universidad Aut\'onoma de Madrid, Cantoblanco 28049 Madrid, Spain}
  \altaffiltext{3}{Department of Astronomy and Astrophysics, Pennsylvania State University, 525 Davey Laboratory, University Park, PA 16802}
  \altaffiltext{4}{Center for Exoplanets and Habitable Worlds, Pennsylvania State University, 525 Davey Laboratory, University Park, PA 16802}

\begin{abstract}

 We report the discovery of a unique object, BD+48 740, a lithium overabundant
 giant with A(Li)=2.33 $\pm$ 0.04 (where A(Li)=$\log n_{Li}/n_{H}+12$),
 that exhibits radial velocity (RV) variations consistent with a 1.6
 M$_{J}$ companion in a highly eccentric, e=0.67 $\pm$ 0.17 and
 extended, a=1.89~AU (P=771~d), orbit. The high eccentricity of
 the planet is uncommon among planetary systems orbiting evolved
 stars and so is the high lithium abundance in a giant star. The
 ingestion by the star of a putative second planet in the system
 originally in a closer orbit, could possibly allow for a
single explanation to these two exceptional facts. If the planet
candidate is confirmed by future RV observations, it might represent the first
example of the remnant of a multiple planetary system possibly affected by stellar evolution.

  \end{abstract}

  \keywords{Stars: fundamental parameters --- Stars: atmospheres --- Stars: late-type --- Stars: individual (BD+48 740) --- Planet-star interactions --- Planets and satellites: detection}


\section{Introduction}
There has been a growing number of exoplanets detected around post-main sequence (MS) stars.
At present, about 50 red giants (RGs) are known to host planetary or brown dwarf-mass companions. 
Recent discoveries also show evidence of multi--planet systems (cf. HD 102272 b, c)
or multi - brown dwarf systems (BD+20 2457 b,c -- \citealt{Niedzielski2009b}).
Although hot jupiter type planets 
exist around sub--giants \citep{Johnson2010},
the more evolved giants show a paucity of short period and eccentric planets \citep{Johnson2007}.
This is most likely the result of tidal disruption and/or planet
engulfment during the RG phase (e.g.\citealt{VillaLivio2009}).

Even rarer is the number of lithium-rich RG stars.
According to the standard evolution theory, when a solar--type star leaves the MS
and becomes a RG, its lithium abundance should drop from A(Li) $\sim 3.3$ to about a 1.5--level.
In fact, the observed upper giant branch limit is A(Li) $<$ 0.5 \citep{LindPrimas2009}
and only a few percent of giants have been observed to have A(Li) $>$ 1.5 (\cite{Kumar2011} and references therein),
with some exhibiting A(Li) $\sim$ 3.3, the value expected for a protostellar disk rather than an evolved star. 

In this paper, we report the discovery of a planet in a highly eccentric, long-period orbit around a lithium-overabundant giant star.

\section{Observations and data analysis}

BD+48 740 is one of the targets of 
the Penn State -- Toru\'n Planet Search (PTPS), which is devoted to
the detection and characterization of planetary systems around stars
more evolved than the Sun.
 
High resolution optical spectra discussed in this paper were collected 
with the Hobby--Eberly Telescope (HET) \citep{Ram1998}. 
The telescope was equipped with the High Resolution Spectrograph (HRS) \citep{Tull1998} which was fed with a 2 arcsec fiber, working at the R = 60\,000 resolution.
For the basic data reduction, standard IRAF\footnote{IRAF is distributed by the National Optical Astronomy Observatories, which are operated by the Association of Universities for Research in Astronomy, Inc., under cooperative agreement with the National Science Foundation.} tasks and scripts were used. 
The signal to noise ratio was typically better than $200--250$ per resolution element at 5900~$\AA$.
For the Li abundance analysis only one order of the red spectra containing the $^{7}$Li 6708$\AA$ line was used.
We have carefully checked all the HET/HRS flat--field spectra for an occasional contamination caused by a nearby feature that, depending on the actual RV of the star, may mimic the Li line and influence the abundance calculation. All such flat-field spectra were excluded from the analysis. 

\section{Results}

BD+48 740 (TYC 3304--90--1) is a V=8.69, 
B-V = 1.252$\pm$0.032, 
$\pi$ = 1.36$\pm$1.13, 
K2 giant in Perseus.
Its basis atmospheric parameters, as well as mass, radius and age were determined by \cite{Zielinski2011} 
 and are presented in Table \ref{tab1}.
 The stellar rotation velocity was estimated by means of the
 \citet{Fekel1997} method and from the SME spectrum modeling. The estimated range of values is given in Table \ref{tab1}. 

\subsection{Radial velocities}
BD+48 740 was observed by the PTPS survey between Jan 12, 2005 and March 5, 2012. 
Precise relative RVs were measured at 15 epochs over 2229 days using the standard I$_2$ cell calibration technique \citep{Butler1996}. 
Doppler shifts were derived from least--squares fits of template
spectra to stellar spectra with the imprinted I$_2$ absorption lines. 

The measured RVs revealed variations over $\sim$ 90 m s$^{-1}$.
With the estimated amplitude of solar--type oscillations
\citep{KjelBedd1995} of $\sim$ 8~ms$^{-1}$ and the average RV uncertainty of 5.6 m s$^{-1}$, 
 the observed RV amplitude was $\sim 10$ times larger than the expected observational uncertainties. Consequently, we found it justified to assume that the observed RV signal could be due to an orbiting planet. 
 
Several absolute RV determinations for BD+48 740 exist in the literature. Three out of five measurements presented in \cite{Abt1970} agree with our determination, while the other two suggest the amplitude of 61~kms$^{-1}$. 
The reason for such a scatter in the RV values is probably the very strong IS Na doublet at $\lambda \lambda$ 5890, 5895 $\AA$, which probably originates in the local IS bubble \citep{Welsh2010}, and is separated from the stellar lines by 46 km s$^{-1}$. When observed at a resolution of 37~$\AA$/mm, the blend shows two peaks with a minimum, whose position depends on the actual atmospheric conditions.
\cite{Famaey2005} found BD+48 740 to show no RV variations, but they computed a different RV for the star. The reason for this discrepancy is not clear.

\begin{table}
\centering
  \caption{\label{tab1}Basic atmospheric and stellar parameters of BD+48 740.}
  
\begin{tabular}{lc}
\hline\hline
Parameter\\
\hline
T$_{eff} [K]$ & 4534 $\pm$24\\
logg & 2.48 $\pm$0.12\\
Fe/H & -0.13 $\pm$ 0.06\\
v$_{micro}$ [km s$^{-1}$] & 1.58\\
v$_{macro}$ [km s$^{-1}$] & 3.2 $\pm$ 0.5\\
 v$_{rot}\sin i$ [km s$^{-1}$] & 3-3.5 \\ 
Vrad [km s$^{-1}$] & -46.32\\
\hline
Log L/L$_{\odot}$ & 1.69 $\pm$0.20\\
M/M$_{\odot}$ & 1.5 $\pm$ 0.3 \\
R/R$_{\odot}$ & 11.4 $\pm$ 0.7 \\
Prot [d] & 165-192 \\
 \hline
 \end{tabular}
 \end{table}
 
\subsection{Radial velocities modeling}
 
\begin{figure}[h] 
  \centering
  \includegraphics[angle=-90, width=0.8\textwidth]{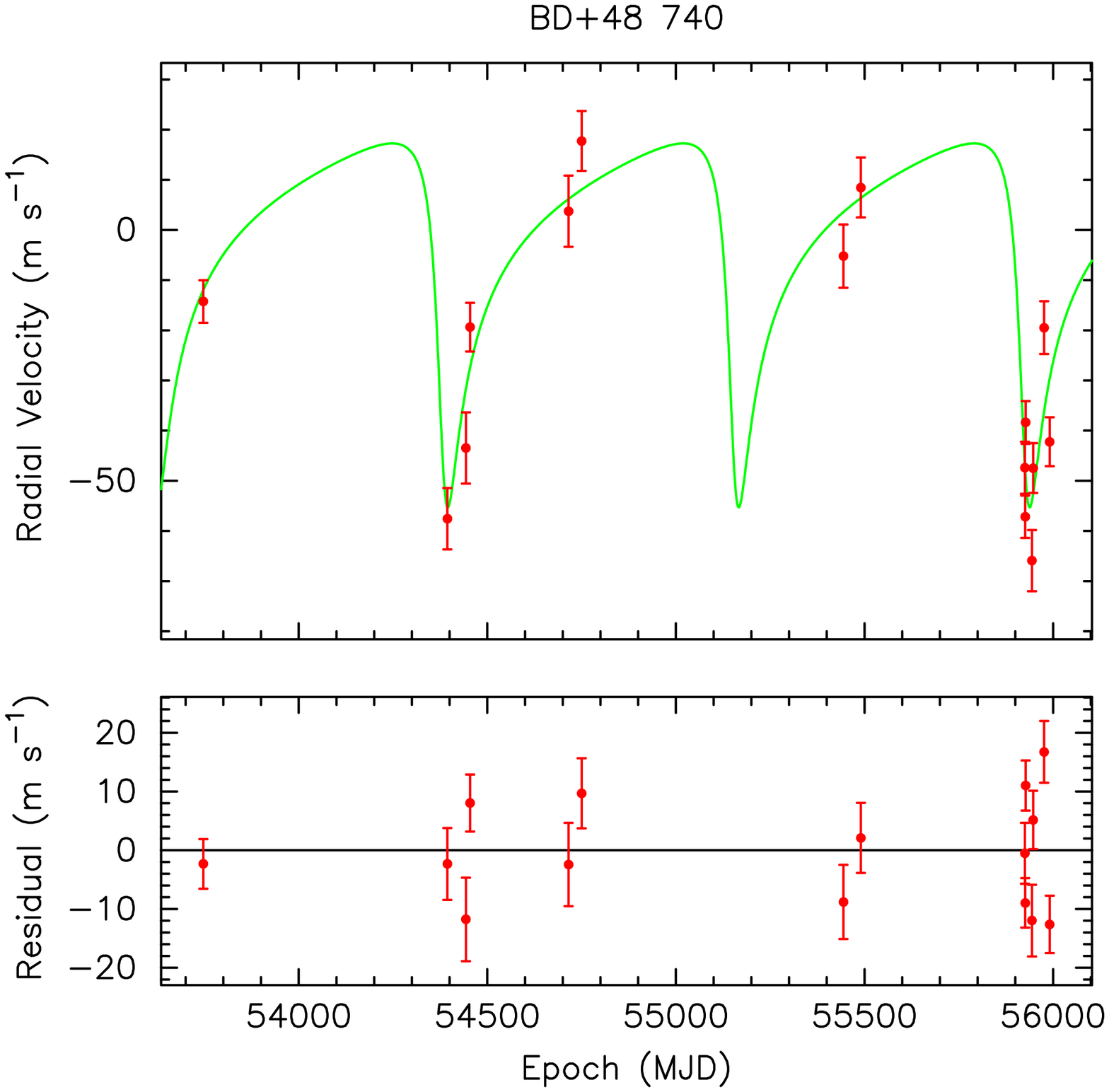} 
  \caption{Preliminary Keplerian fit to 15 epochs of RV observations of BD+48 740.}
  \label{fig1}
\end{figure}

{The initial, semi--global search of the orbital parameter space to model the observed RV variations was carried out with the PIKAIA code \citep{Charb1995}. This was followed by the Levenberg--Marquardt, least-squares modeling of the RVs in terms of the standard, six--parameter Keplerian orbit 
within a narrow range selected by the genetic algorithm.} All the available measurements were used and no stellar jitter was added to data. The resulting orbital solution is presented in Table \ref{tab2} and in Figure \ref{fig1}. 
Uncertainties in the best-fit orbital parameters (Table \ref{tab2}) were estimated from the parameter covariance matrix.

 \begin{table}
 \caption{\label{tab2}Orbital solution for BD+48 740 b.}
\centering
\begin{tabular}{lc}
\hline\hline
Parameters\\
\hline
P [days] & 771.3$\pm$ 7.4\\
T$_0$ & 55925.32$\pm$ 6.8\\
e & 0.67 $\pm$ 0.17 \\
K [m s$^{-1}$]& 36.3 $\pm$ 9.1\\
$\omega$ & 140.5$\pm$ 47.3\\
\hline
$m_2\sin (i) [M_{Jup}]$ & 1.6 \\
a [AU] & 1.89 \\
\hline
$\chi ^2$ & 4.86 \\
 $\sigma_{RV} $  [m s$^{-1}$] & 9.26\\
 \hline
 \end{tabular}
 \end{table}

To check the reliability of the preliminary orbit, 
1000 trials of scrambled radial velocities 
were performed, which resulted in 112 solutions with $\chi^2$ lower than the one presented here. Therefore, the corresponding false alert probability for our best--fit model is FAP=11.2$\%$.

The resulting post--fit residuals are consistent with stellar jitter interpreted as under--sampled p--mode oscillations. 
No correlation between the observed radial velocity and the line bisector variations exists (r=0.166) and the correlation between radial velocities and line profile curvature (r=0.488) is statistically insignificant. 

The fact that the orbital period is 
much longer than the estimated rotation period, 
and the high orbital eccentricity make it unlikely that the RV variations are generated by a spot rotating with the star. 

We have also searched 3029 WASP 
photometric measurements of BD+48 740 collected over 1475 days for any periodic variations and found none within uncertainties.
Furthermore, 134 epochs of Hipparcos photometry of the star reveal its constant brightness of H$_p$=8.8543 $\pm$ 0.0017 mag. 

Therefore, we conclude that the observed RV variations reflect a Keplerian motion of the star around the system's barycenter.

 \subsection{Li abundance }

The observed spectra were analyzed with the Spectroscopy Made Easy package (SME, \citealt{SME1996}). SME assumes LTE and parallel plane geometry, but it ignores magnetic fields, molecular lines and mass loss. 
The synthetic spectrum synthesis was based on Kurucz's models of stellar atmospheres.

 \begin{figure}[h] 
  \centering
  \includegraphics[angle=0, width=0.8\textwidth]{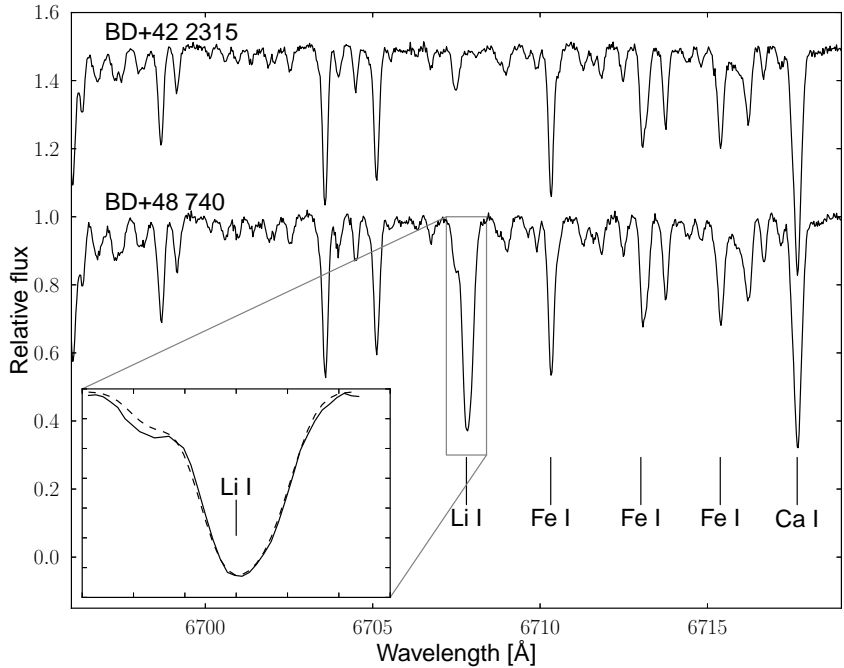} 
  \caption{Observed HET/HRS spectra (continuous line) of BD+48 740 and BD+42 2315, a low lithium giant with A(Li)=0.23, in a narrow range centered at Li I 6708~$\AA$. The final SME fit to the spectrum of BD+48 740 for the Li line is presented in the inset with the dashed line. }
  \label{fig2}
\end{figure}

As an input, SME requires a set of lines from the Vienna Atomic Line Database (VALD, \citealt{VALD1999}) identified in the spectrum, the stellar data (including RV), and the instrumental profile.
The observed spectrum was considered as a model constraint. 
Although SME requires the M/H ratio, the
Fe/H was used as an approximation for the metallicity.
The 6695-6725~$\AA$ range was modeled,
which covers the $^7$Li line at 6708~$\AA$ as well as 
several lines of Al, Ti, Si and Ca -- Fig.\ref{fig2}. SME was allowed to fit RV, rotation velocity 
and macroturbulence velocity as well. 
To increase the precision, SME fitting was performed on all 16 spectra obtained with the HET, 
yielding an estimate of the abundance uncertainty (rms).
A comprehensive analysis of the lithium abundances also requires a
correction for non--LTE effects, especially in the case of the 
lithium line at 6708 $\AA$ considered here, which is
affected by several non-LTE processes 
(\cite{Carlsson1994} and references therein). 
With SME LTE A(Li)=2.24, and the non--LTE correction of 0.087 from \cite{Lind2009}, the final non--LTE value of A(Li)=2.33 $\pm$ 0.04 (rms) was obtained.
This value places BD+48 740 on the short list of lithium overabundant giants.

\section{Discussion}

The effective temperature (known quite precisely) allows the placement
of BD+48 740 close to the stable helium
burning part of the evolutionary track of a 1.5 M$_{\odot}$, solar
metallicity star of
\cite{Girardi2000}. Although a relatively small spread in the evolutionary tracks in this region of the
HR diagram, together with the uncertainty in the luminosity, does not allow an unambiguous determination of BD+48 740
evolutionary status, the location of the star on the HR diagram favors the interpretation in which
the star is evolving toward the Red Giant Branch (RBG) tip. However, we cannot completely exclude
scenarios in which the star is descending from the giant branch toward the horizontal branch (HB)
or ascending toward the AGB. 
Unfortunately our HET/HRS spectra do not allow for the determination of the
the $^{12}$C/$^{13}$C carbon isotopic ratio to constrain the evolutionary status better.

Any interpretation of the current state of BD+48 740
past the RGB tip would imply that the star had already gone through the
helium flash and extended envelope stage. 
As a result, the observed planet at 1.9~AU would have been disrupted by
tidal capture or at least have its orbit circularized
\citep{VillaLivio2009}. 
Given the star's evolutionary status and the orbital distance of the planet,
orbital circularization is not expected for another 0.5 Gyr.

\subsection{Eccentricity}

The observed orbital eccentricity may be overestimated due to the small number of data points available
\citep{Zakamska2011} or it could result from not resolving two components in 2:1 circular or near-circular resonant orbits
\citep{Ang-Esc2010}. If taken as a true value, the observed eccentricity of BD+48 740 b at 1.89 AU, although high,
fits into the observed distribution of the RV planets as a function of 
semi--major axis (e.g. \citealt{Kane2012}). This distribution begins to significantly 
diverge from purely circular orbits beyond 0.04 AU, and by 0.1 AU it reaches the eccentricity range of 0.0 -- 0.5
\citep{Kane2012}. It is interesting to note that the
distribution of eccentricities in multiple planet systems differs from that of single planets \citep{Wright2009b} in that it displays lower
values. Moreover, when plotted as a function of log (g) of the host star, the distribution of orbital eccentricities of planets orbiting 
giants narrows, showing generally mildly eccentric orbits, typically with e$<$0.45 with only one exception of HD 1690 (e=0.64) \citep{Moutou2011}.

Since tidal interaction of the planet with BD+48 740 is still not
expected at this point and the observed eccentricity is not unusual among planets orbiting MS
stars, we could argue that the planet eccentricity is {\it
primordial}, that is, it has not been modified as the star 
left the MS. However, if we consider the fact that the eccentricity is very high for a giant star companion
we then need to find a mechanism operating only during the post--MS evolution of the star
that resulted in such eccentricity. 

In principle, post--MS evolution can trigger instability in a multiple system that was stable during the stellar
MS evolution due to stellar mass-loss
\citep{DebSig2002}. However, RGB stars are not expected to be losing
mass in large quantities and therefore the stability boundary should
not move considerably during this phase. Moreover, a non--adiabatic
approximation is neither expected to operate at the current planet orbital
distance and stellar mass-loss regime \citep{VillaLivio2007} nor to excite
the eccentricity of the planet.

It is generally accepted that large eccentricities, at any
evolutionary stage, have to be excited (see
e.g. \citealt{FordRasio2008}). Of the several mechanisms that have
been proposed, we find the Kozai effect \citep{TakRas2005} very unlikely to take
place in BD+48 740 b, since no binary companion
has been detected with a good degree of confidence. 
The nearest object to BD+48 740 is 2MASS 02425927+4855549, a J=13.288 mag star which is separated
by 12.3 arcsec, corresponding to $\sim$6840 AU (assuming the
distance to BD+48 740 d=556 $\pm$ 129 pc
from 
\cite{Famaey2005}). The JHK photometry available for the star
suggests however, that the 2MASS object is a hot, distant field
star. Furthermore, the star is not a member of the nearby $h$ and $\chi$ Per
open cluster, and its proper motion show that it is not a high velocity star,
thus making it very unlikely that a stellar encounter would be responsible
for exciting the eccentricity \citep{LaughlinAdams1998}. 

We are left then with a planet--planet scattering event as a plausible
mechanism to excite the eccentricity of this evolved system (see
e.g. \citealt{FordRasio2008}), either during the MS or during the
post--MS evolution of the star. A viable scenario could be one in
which the the BD+48 740 planetary system contained originally more than one planet.
In turn of a planet-planet scattering event, the
innermost (unobserved) planet was brought
into the stellar surface due to either a direct hit or tidal
interaction with the evolving star, while the orbit of the outer planet was excited to the observed, high eccentricity
\citep{Marzari2002, Mardling2010}.

The derived stellar parameters of BD+48 740 imply a stellar radius of R/R$_{\odot}$ = 11.4 $\pm$ 0.7 (or 0.05 AU).
A small number of confirmed exoplanets around MS stars orbit within this orbital distance of their host
stars (see e.g. \citealt{CollierCameron2007}).
It is very likely that
these planets, at such small orbital separations, will tidally interact with their host already
while on the MS with the end result of the spiral-in of the planet into the stellar
envelope (see e.g. \citealt{Jackson2009}). More substantially, for Sun like stars,
 the number of Jovian-type planets in the hot jupiter
category, orbiting within 0.1~AU, is about 1\% \citep{Mayor2011}.
It is expected, and indeed confirmed by the occurrence fraction of hot jupiters around sub--giant
stars with masses M$_*>$ 1.45 M$_{\sun}$ \citep{Johnson2010}, that a large majority of these
systems, with the exception of a few extreme cases, will survive the MS evolution. However, after the star leaves the MS, and especially as it ascends the RGB, the tidal influence
becomes important and extends far beyond the stellar radius for these almost fully convective stars \citep{VillaLivio2009}.

The BD+48 740 planetary system might have originally resembled HAT
P-13 b, c \citep{Bakos2009}
or HD 217107 b, c \citep{Fischer1999}. 
 These are the systems around MS stars,
both of which have a close--in planet in a nearly circular orbit (a hot jupiter) and another,
more distant companion in an eccentric orbit. 
A possible
dynamical evolution leading to such configuration, involving
planet-planet scattering, was presented by \cite{Mardling2010}.
The observed enhanced Li abundance suggests, however, that in the case of
BD+48 740 the planet-planet scattering event operated quite
recently, during the post-MS evolution of the star.

\subsection{Lithium overabundance}

It is difficult to identify a mechanism that might have preserved the primordial Li-abundance of BD+48 740
given that, as soon as the the star moved past the sub-giant branch, convection should have reduced its Li
abundance by at least an order of magnitude (see e.g. \citealt{GrattonSneden2000}). 

Thus, either BD+48 740 has experienced a Li-production in its interior, or, it has suffered external pollution from a companion. We can rule out the presence of a stellar companion that might have transfered Li-rich material onto the surface of BD+48 740, since the RV measurements do not show any sign of it. 

Lithium could have been produced in the interior of BD+48 740, brought up to the stellar surface and survived, if a Cameron-Fowler mechanism \citep{CamFow1971} had been in place in this low mass RGB star. 
For that, high mixing rates between the surface and the Li-forming regions are required. It has been shown by several works now that Li-rich stars can be found all along the RGB (see e.g. \citealt{
LebUtt2012}) thus excluding the \cite{CharBal2000} process, in which the extra mixing, and therefore the Li enrichment, should be associated with the RGB bump, when the molecular weight discontinuity created by the inward penetration of the convective envelope is erased by the advancement of the H-burning shell. Even allowing for uncertainties, the location of BD+48 740 on the HR diagram
supports the conclusions of these works. 

Several other sources for extra mixing have been suggested to explain
Li--rich giants, i.e. tidally enforced enhanced extra mixing
\citep{DenHer2004}, thermohaline \citep{CharZahn2007},
and Magneto-Thermohaline Mixing \citep{DenPinMac2009}. In fact, they could be operating all together in
BD+48 740, if we assume a scenario in which a
companion has been tidally
captured and engulfed into the stellar surface (see \citealt{VillaLivio2009}). In particular, it has been recently shown by \cite{Garaud2011} and
\cite{TheadoVauclair2012} that the accretion of metal rich planetary
material onto the surface of exoplanet host stars can increase the Li
surface abundance of the target stars on timescales that depend on the
mass and stellar structure and have been estimated to be short,
of the order of a few Myr (see also
\citealt{DenHer2004}). \cite{SiessLivio1999} have shown that
engulfment of substellar objects 
could also trigger Li
enhancement and a mass-loss enhacement on the
RGB. No trace of IR excess can be found in the 2MASS and WISE 
data for BD+48 740 (J-K vs. K-WISE[3.4, 4.6, 12 and 22 $\mu$m]). 
Archive IRAS observations, upper limits mostly, suggest however, that the 
star is in the extended shell and relatively high lithium abundance phase of \cite{SiessLivio1999} scenario.

The projected rotation velocity of BD+48 740 is typical for its luminosity
and effective 
temperature. The actual amount of stellar rotation velocity increase depends
strongly on stellar structure (envelope mass) and the
(unknown) rate of planet evaporation. The observed rotation
velocity of BD+48 740 neither contradicts nor supports the engulfment scenario.

\section{Conclusions}
We present evidence of a high lithium abundance, A(Li)=2.33 $\pm$
0.04, in the RG star BD+48 740. This value is well above the expected limit of
A(Li)=1.5 for evolved stars. We also present multi-epoch, precise radial velocities for
the star which, although sparse, show periodic variations which can
be interpreted as a result of the Keplerian motion of a planetary
mass companion with $m\sin i=1.6$ M$_{J}$, on a highly eccentric orbit
of e=0.67. We find no evidence of a stellar-mass companion to BD+48 740. 

Given the current evolutionary status of the star, its Li abundance
and the planet's current orbit, we discuss a possibility that BD+48 740
had a second planet in an innermost orbit that could have been engulfed
by the star. This possibility, although not directly verifiable
with the available data, allows a scenario in which both the high
lithium abundance of the star and the planet's highly eccentric orbit could originate
from a recent, violent dynamical event, likely representing a later stage in the evolution of the planetary system,
which was similar to HAT P-13 while still on the MS.
 
BD+48 740 is the first example of Li-overabundant giant star with a
planetary mass companion 
candidate. 

\section{Acknowledgements}
We thank Dr. Nikolai Piskunov for making SME available for us. We
thank the HET resident astronomers and telescope operators for continuous 
support. MA, AN and GN were supported by the Polish Ministry
of Science and Higher Education grant N N203 510938. AW was supported
by the NASA grant NNX09AB36G. 
EV acknowledges the support provided by the Spanish Ministry of
Science and Innovation (MICINN) under grant AYA2010-20630 and to the 
Marie Curie FP7-People-RG268111. The HET is a joint project of the
University of Texas at Austin, the Pennsylvania State University,
Stanford University, Ludwig-Maximilians-Universit\"at M\"unchen, and
Georg-August-Universit\"at G\"ottingen. The HET is named in honor of
its principal benefactors, William P. Hobby and Robert E. Eberly. The
Center for Exoplanets and Habitable Worlds is supported by the
Pennsylvania State University, the Eberly College of Science, and the
Pennsylvania Space Grant Consortium. 
This research has made extensive
use of the SIMBAD database, operated at CDS (Strasbourg, France) and
NASA's Astrophysics Data System Bibliographic Services. 
This research has made use of the Exoplanet Orbit Database
and the Exoplanet Data Explorer at exoplanets.org.

\bibliographystyle{apj} 

\end{document}